\RequirePackage{xcolor} %
\documentclass[journal,twoside,web]{ieeecolor}
\usepackage{tmi}
\usepackage{amsmath,amssymb,amsfonts}
\usepackage{graphicx}
\usepackage{textcomp}
\usepackage[
    style=ieee,            %
    doi=false,             %
    isbn=false,            %
    url=false,              %
    eprint=false,          %
    maxbibnames=6,         %
    minbibnames=1,         %
    giveninits=true,       %
    sorting=none,          %
    backend=biber
]{biblatex}

\usepackage{jabbrv}
\DeclareFieldInputHandler{journaltitle}{%
  \def\NewValue{\JournalTitle{#1}}}
 
\makeatletter
\DeclareFieldInputHandler{booktitle}{%
  \iffieldequalstr{entrytype}{inproceedings}
    {\def\NewValue{\JournalTitle{#1}}}
    {\def\NewValue{#1}}}
\makeatother

\def\BibTeX{{\rm B\kern-.05em{\sc i\kern-.025em b}\kern-.08em
    T\kern-.1667em\lower.7ex\hbox{E}\kern-.125emX}}

\definecolor{wessel}{HTML}{129f57}
\definecolor{darkorange}{HTML}{9d5600}

\newcommand{\wn}[1]{{\color{black}#1}}
\newcommand{\on}[1]{{\color{black}#1}}

\newcommand{\wes}[1]{{\color{black}#1}}
\newcommand{\west}[1]{\textcolor{black}{#1}}
\newcommand{\ois}[1]{{\color{black}#1}}
\newcommand{\wtwo}[1]{{\color{black}#1}}
\newcommand{\otwo}[1]{{\color{black}#1}}
\newcommand{\othree}[1]{{\color{black}#1}}
\newcommand{\wthree}[1]{{\color{black}#1}}
\newcommand{\moved}[1]{#1}

\newcommand{\strike}[1]{} %

\usepackage{etoolbox}
\makeatletter
\@ifundefined{color@begingroup}%
  {\let\color@begingroup\relax
   \let\color@endgroup\relax}{}%
\def\fix@ieeecolor@hbox#1{%
  \hbox{\color@begingroup#1\color@endgroup}}
\patchcmd\@makecaption{\hbox}{\fix@ieeecolor@hbox}{}{\FAILED}
\patchcmd\@makecaption{\hbox}{\fix@ieeecolor@hbox}{}{\FAILED}

\usepackage[nolist,nohyperlinks,printonlyreused]{acronym}
\begin{acronym}
\acro{DM}{diffusion model}
\acro{MRI}{magnetic resonance imaging}
\acro{CT}{computed tomography}
\acro{CS}{compressed sensing}
\acro{DL}{deep learning}
\acro{ELBO}{evidence lower bound}
\acro{SMC}{sequential Monte Carlo}
\acro{PF}{particle filter}
\acro{VAE}{variational autoencoder}
\acro{DNN}{deep neural network}
\acro{BCE}{binary cross entropy}
\acro{SGD}{stochastic gradient descent}
\acro{TPR}{true positive rate}
\acro{FPR}{false positive rate}
\acro{CNN}{convolutional neural network}
\acro{MI}{mutual information}
\acro{ViT}{visual transformer}
\acro{IoU}{intersection over union}
\acro{IIC}{invariant information clustering}
\acro{FCN}{fully convolutional network}
\acro{PhC}{phase contrast}
\acro{DIC}{differential interference contrast}
\acro{EM}{electron microscopy}
\acro{GAN}{generative adversarial network}
\acro{ROIs}{regions of interest}
\acro{ROI}{region of interest}
\acro{TIMM}{PyTorch image models}
\acro{AWGN}{additive white Gaussian noise}
\acro{PCA}{principal component analysis}
\acro{SSL}{self-supervised learning}
\acro{SAM}{segment anything model}
\acro{C.I.}{confidence interval}
\acro{RBF}{radial basis function}
\acro{DPS}{diffusion posterior sampling}
\acro{PSNR}{peak signal-to-noise ratio}
\acro{DICE}{Dice-Sørensen coefficient}
\acro{RTBF}{retrospective transmit beamforming}
\acro{gCNR}{generalized contrast-to-noise ratio}
\acro{FWHM}{full-width at half-maximum}
\acro{LPIPS}{learned perceptual image patch similarity}
\acro{FOV}{field of view}
\acro{SNR}{signal-to-noise ratio}
\acro{MLT}{multi-line transmission}
\acro{SAMI}{simultaneous axial multifocal imaging}
\acro{PRF}{pulse-repetition frequency}
\acro{HFR}{high frame-rate}
\acro{CASL}{Cognitive ultrasound Acquisition of Scan Lines}
\end{acronym}

\newcommand{\bh}{\bm{h}}

\newcommand{\bn}{\bm{n}}
\newcommand{\bx}{\bm{x}}
\newcommand{\by}{\bm{y}}

\newcommand{\bep}{\bm{\epsilon}}

\newcommand{\sfi}{\mathsf{i}}

\newcommand{\tens}[1]{\bm{\mathsf{#1}}}
\newcommand{\tA}{\tens{A}}

\newcommand{\tY}{\tens{Y}}
\newcommand{\tX}{\tens{X}}

\DeclareMathOperator*{\argmax}{arg\,max}

\newcommand{\circled}[1]{%
  \tcbox[
    colback=white,                 %
    colframe=black,                %
    boxrule=0.4pt,                 %
    arc=5pt,                     %
    outer arc=5pt,               %
    boxsep=0pt,                    %
    left=2.5pt, right=2.5pt,       %
    top=1.5pt, bottom=1.5pt,       %
    nobeforeafter,                 %
    halign=center,                 %
    valign=center,                 %
  ]{\textbf{#1}}                 %
}

\usepackage[hidelinks]{hyperref}
\usepackage{booktabs}
\usepackage{algorithm}
\usepackage[noend]{algpseudocode}
\usepackage{tcolorbox} 
\usepackage{setspace}
\usepackage{bm}
\usepackage{soul} %

\begin{document}
\bstctlcite{IEEEexample:BSTcontrol}
\title{Patient-Adaptive Echocardiography using Cognitive Ultrasound
{
\thanks{
Manuscript received 12 August 2025; revised 23 January 2026; accepted 28 April 2026.
This work was supported by the European Research Council (ERC) under the ERC starting grant nr. 101077368 (US-ACT). Wessel L. van Nierop and Oisín Nolan contributed equally to this work. Wessel L. van Nierop, Oisín Nolan, Tristan S.W. Stevens, and Ruud J.G. van Sloun are with the Department of Electrical Engineering, Eindhoven University of Technology, 5612 AZ Eindhoven, The Netherlands (email: w.l.v.nierop@tue.nl; o.i.nolan@tue.nl; t.s.w.stevens@tue.nl; r.j.g.v.sloun@tue.nl)}
}
}
\author{Wessel L. van Nierop, \IEEEmembership{Member, IEEE}, Oisín Nolan, \IEEEmembership{Member, IEEE}, \\
Tristan S.W. Stevens, \IEEEmembership{Member, IEEE}, and Ruud J.G. van Sloun, \IEEEmembership{Member, IEEE}
}

\maketitle

\begin{abstract}
Focused transmits are the most commonly used \wes{transmit strategy} for echocardiograms, but suffer from relatively low frame rates, and in 3D, even lower volume rates. Fast imaging based on unfocused transmits \on{has} disadvantages such as motion decorrelation and limited harmonic imaging capabilities. This work introduces a patient-adaptive focused transmit \wes{and receive} scheme that has the ability to drastically reduce the number of transmits needed to produce a high-quality ultrasound image. The method relies on posterior sampling with a temporal diffusion model to perceive and reconstruct the anatomy based on partial observations, while subsequently acquiring the most informative transmits. This \wes{cognitive ultrasound} modality outperforms random and equispaced subsampling \wes{in terms of distortion and perceptual metrics} on the 2D EchoNet-Dynamic dataset and a 3D Philips dataset, where we actively select focused elevation planes. 
Furthermore, our method \wtwo{improves \acl{gCNR} from 0.83 to 0.89} compared to the same number of diverging wave transmits on six in-house echocardiograms.
Additionally, we can \wes{segment the left ventricle, with on average 0.91 \acl{DICE},} \wtwo{through simulating using 2 out of 112 lines.}
Finally, our method can be run in real-time on GPU accelerators from 2023, \wes{increasing the maximum achievable frame-rate \wtwo{from 46 Hz to 58 Hz}.} \wn{The code is publicly available at \underline{\href{https://tue-bmd.github.io/casl/}{https://tue-bmd.github.io/casl/}}.}
\end{abstract}

\acresetall

\section{Introduction}
\label{sec:introduction}

\begin{figure}
    \centering
    \includegraphics[width=\linewidth]{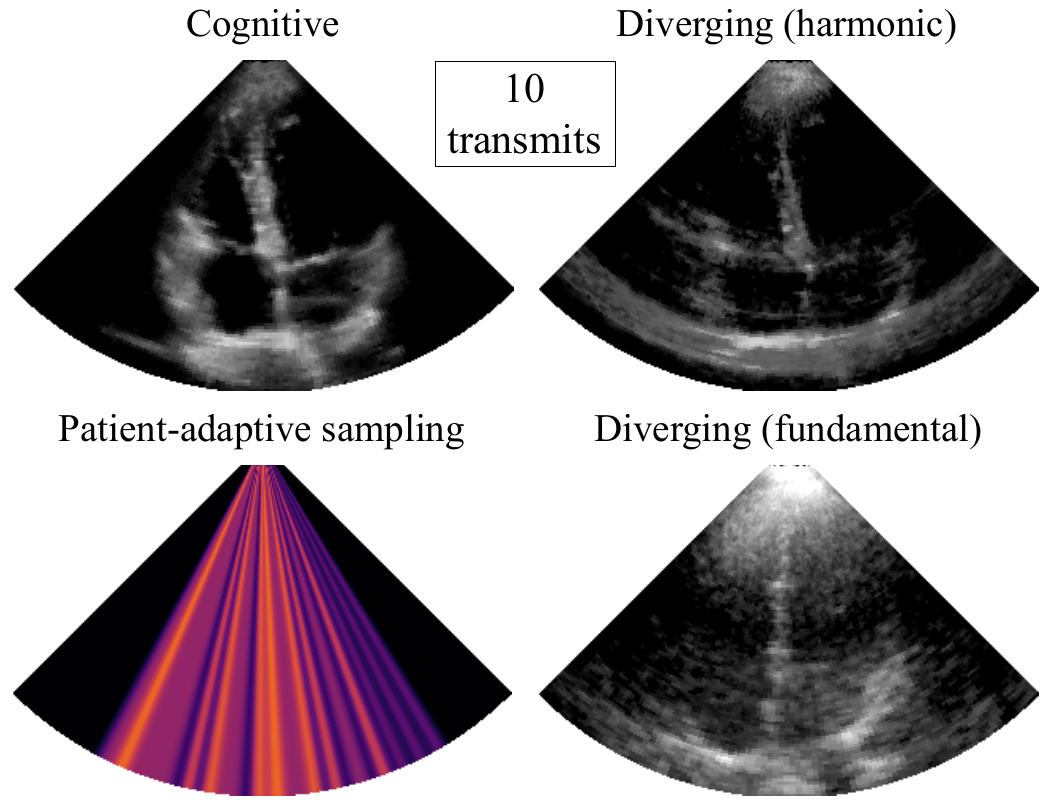}
    \vspace{-9pt}
    \caption{\wes{\ac*{CASL} reduces the number of acquisitions needed to obtain a high-quality ultrasound image by actively selecting those measurements that are expected to be most informative. Each image is created using 10 transmit events.}}
    \label{fig:teaser}
\end{figure}

\IEEEPARstart{U}{ltrasound} imaging is one of \wn{the} most widely used medical imaging modalities. It \wtwo{offers} advantages that other modalities such as \ac{MRI} and \ac{CT} do not bring, such as being affordable, portable, real-time, and non-ionizing. These advantages make ultrasound very accessible~\cite{wise_everyones_2008}.

\wtwo{Among its clinical applications, echocardiography is the first-line non-invasive technique for assessing cardiac structure and function. It is essential for diagnosing and monitoring major cardiovascular conditions, including heart failure, heart valve disease, cardiomyopathy, ischemic heart disease, and pulmonary hypertension~\cite{lang_recommendations_2015}.}

\wes{
There exist various techniques for ultrasound image formation.
\wtwo{The transmit scheme can be designed using focused or unfocused events. Focusing permits the concentration of acoustic energy} at specific locations within the body.
Focusing the beam will provide enhanced lateral resolution, \ac{SNR}, and penetration depth in echocardiography~\cite{kakkad_effect_2018}.
Another important reason to opt for focused transmits is the generation of high-amplitude pressure fields, which are necessary for the generation of harmonic components used in harmonic imaging~\cite{demi_parallel_2012}. Harmonic imaging has become the gold standard for echocardiograms due to the superior image quality in hard-to-image patients, reducing reverberation artifacts, clutter, and near-field artifacts~\cite{thomas_tissue_1998,hawkins_original_2008, monaghan_second_2000,matte_optimization_2011, schmitz_ultrasound_2020, kakkad_effect_2018}. 
Unfocused transmit events, such as diverging waves, are typically coherently compounded to improve \ac{SNR} \cite{montaldo_coherent_2009}. However, compounding can result in motion decorrelation in high-motion scenarios, which means that the tissue has moved a significant amount in between the transmit events, reducing image quality \cite{grondin_cardiac_2017,nie_high-frame-rate_2019,nie_motion_2019}.
Due to these advantages, focused transmits are the most widely used transmit strategy in commercial ultrasound systems~\cite{vansloun2021deeplearningultrasoundbeamforming}. The downside is that the excited area is much smaller, meaning that more transmit events are necessary to cover a given \ac{FOV} compared to unfocused transmit events, reducing the temporal resolution. 

\wtwo{For accurate diagnosis in echocardiography a frame rate of at least 40 Hz is required \cite{NEGOITA201631}, while for certain medical conditions, such as tachycardia, over 80 Hz is advisable \cite{mor_avi_current_2011}.}
Temporal resolution in ultrasound image formation is determined by the number of transmit events, imaging depth, and the speed of sound. For an imaging depth of 15~cm and a typical sound speed of 1540~m/s (both common in echocardiography), each transmit requires 195~$\mu$s, which translates to a frame rate of over 5~kHz. Because focused transmits do not cover a wide \ac{FOV} with high \ac{SNR} and high-amplitude pressure fields, many transmit events are often necessary. 
\wtwo{For example, a phased-array transducer with an aperture of 20 mm and center frequency of 3 MHz, has an angular resolution of approximately 1.5\textdegree, which means 122 lines are needed for the full angular resolution (half beamwidth spacing) in a 90\textdegree\text{ }sector \cite{von_ramm_beam_1983}. This reduces the best achievable frame rate to 42~Hz, or 21~Hz when using pulse inversion for harmonic imaging \cite{shen_pulse_2005}.} \ois{In order to image phenomena with higher temporal resolution, for example valve dynamics, one might narrow the field of view or resort to M-mode imaging \cite{zhong2021visualization}. This, however, imposes a trade-off between field of view and temporal resolution.} For 3D echocardiograms, even more focused transmit events are \ois{necessary}, meaning that it is} hard to obtain high-quality and fast 3D \wes{echocardiograms}. This shows a need for a reduction in the number of transmit events while maintaining high image quality to ensure diagnostic accuracy.

In addition to accelerating frame rates, reducing the number of necessary transmit events also reduces certain cost factors associated with the acquisition. \ois{Such cost factors include power usage and on-device compute, which currently bottleneck} imaging modalities that depend on battery power and require small form factor, such as wearable ultrasound patches for continuous monitoring \cite{huang2023emerging, ottakath2024wearable}. Another cost factor is the bandwidth required to communicate the acquired data to a server for processing, which is of particular relevance to cloud-based ultrasound \cite{hadri2024ultrasound}.

This work aims to reduce the number of acquisitions needed to obtain a high-quality ultrasound image by actively selecting those measurements that are expected to be most informative \ois{(Fig. \ref{fig:teaser})}. This fits into the paradigm of \textit{cognitive ultrasound}, \ois{recently proposed by van Sloun~\cite{van_sloun_active_2024}, in which the imaging process is modeled as an autonomous agent that actively designs future transmit events to maximize information gain. We achieve this by equipping an imaging agent with a generative model of the ultrasound scene and observations, tracking beliefs about plausible anatomical explanations for the measurements it observes. Based on these beliefs, the agent pursues \wes{focused} acquisitions that have the highest expected information gain. \otwo{This results in a drastic reduction in} the number of \wes{focused} transmit events per frame and thus increases the frame rate, \ois{offering} a potential alternative to unfocused transmits. \otwo{We refer to our method throughout the paper as \textit{CASL}: \textit{\textbf{C}ognitive ultrasound \textbf{A}cquisition of \textbf{S}can \textbf{L}ines}.}}

\subsection{Related Work}
\otwo{Reducing the number of transmit events necessary to create an ultrasound image has the effect of reducing the data rate and increasing the maximum potential frame rate. In this section, we review a number of existing approaches to achieving these goals. 

Firstly, we consider methods that optimize the focusing strategy to gain more information about the target per transmit. \wes{\Ac{MLT}, for example, generates multiple focal points during transmission, such that the number of transmit events is reduced. However, the parallel focused beams do suffer from interference, which can result in visible artifacts in the image~\cite{demi_practical_2018}. Furthermore, methods such as \ac{SAMI}~\cite{ilovitsh_simultaneous_2019}, improve the level of focusing along the axial dimension, decreasing the need for focusing along different depths using multiple transmits. \Ac{MLT} and \ac{SAMI} share the same goal as this paper, but are complementary to our method, which can use any focused transmit.}} 

\othree{Çakiroğlu \textit{et al.} \cite{ccakirouglu2023autoencoder} take an alternative approach, optimizing a single set (non-adaptive) of transmit delays and weights through backpropagation to improve the recovery of imaging parameters across a large dataset of simulated examples. CASL, in contrast, takes a cognitive approach, optimizing transmit parameters adaptively.}

\othree{Another popular approach to reducing data rates has been to use \ac{CS} \cite{donoho2006compressed}.} For instance, \othree{Chernyakova \textit{et al.} \cite{chernyakova2014fourier} propose a Fourier-domain formulation, recognizing that the Fourier coefficients of the received signals can be obtained from their low-rate samples. \ac{CS} has also been applied to recover fully-sampled RF channel data from subsampled measurements (e.g. RF samples or entire channels/elements), for example by leveraging sparsity in the wave atom basis ~\cite{ramkumar2019strategic, friboulet2010compressive}.} Beyond subsampling, other work has proposed compressive multiplexing of channels \cite{besson2018compressive, besson2018joint}. \othree{For reconstruction beyond conventional sparsity priors, \ac{DL} \cite{goodfellow2016deep} has become increasingly attractive. In ultrasound, it has been used to recover beamformed scan lines and fully-sampled RF channel data from e.g. sparse array designs \cite{mamistvalov2022deep, xiao2022minimizing}. As with MLT and SAMI, such approaches to reducing the channel data rates can also be used in combination with CASL, which reduces the number of transmit events (transmit rates).}

\otwo{Other methods subsample at the level of the transmit, as CASL does. For example, the approach by Huijben \textit{et al.}~\cite{huijben2020learning} learns subsampling masks for both channels and slow-time frames jointly with a \ac{CNN} reconstruction model, using the Gumbel-Max trick~\cite{huijben2022review} to backpropagate through the subsampling operation. Similarly, Lorintiu \textit{et al.} \cite{lorintiu2015compressed} employ dictionary learning to reconstruct 3D ultrasound volumes from subsampled scan lines. Afrakteh \textit{et al.} also tackle scan-line subsampling, using random masks with tensor completion for reconstruction ~\cite{afrakhteh2022high}. Each of the above methods uses either a fixed mask, optimized for a particular task or group of patients, or a random mask. CASL, in contrast, designs \textit{patient-adaptive} sampling masks, in real-time. Given the variability that exists across patient anatomies, patient-adaptive subsampling is essential to minimize redundancy and therefore maximize information gain.} While patient-adaptive subsampling algorithms have been proposed for other medical imaging modalities, such as \ac{MRI}~\cite{van2021active, yin2021end, nolan2025active, yiasemis2024end} and X-Ray \ac{CT} \cite{wang2023active}, this work is, to our knowledge, the first application of patient-adaptive subsampling to ultrasound video recovery.

In this work, we identify the task of recovering fully-sampled ultrasound frames from a subset of scanned lines as being akin to \textit{inpainting}, a popular task in computer vision and image generation: in both cases, the goal is to recover the missing portion of the signal. We therefore choose to use diffusion models, which have shown excellent performance in inpainting~\cite{chung2022diffusion, rout2024solving} to solve this problem, outperforming traditional \ac{CS} and supervised \ac{DL} methods \cite{song2022inverse}. This modeling choice is further motivated by recent success in applying diffusion models to the domain of ultrasound, for synthetic data generation~\cite{stojanovski2023echo}, dehazing~\cite{stevens2024dehazing}, and \othree{image reconstruction}~\cite{zhang2023ultrasound}.

\subsection{Contributions}
\otwo{To summarize, t}his paper presents the following main contributions. (1)~We propose a method for reconstructing ultrasound video using a minimal number of transmit events with a temporal diffusion model that exploits the sequential nature of ultrasound. (2) We propose an algorithm for designing a transmit scheme which maximizes information gain in \on{a computationally efficient way}. (3) \otwo{We present experimental results showing} that acquiring patient-adaptive focused transmits outperforms diverging waves for the same number of transmit events in terms of \ac{gCNR}.

\section{Theoretical Background}

\subsection{Active Perception}
\label{sec:active-perception}
The goal of sensing is to acquire measurements to gain information about parameters describing the state of some environment of interest. Often, however, the acquisition process has some constraints -- for example, a limited field of view might require that the sensor is steered in order to capture a certain aspect of the environment~\cite{bajcsy1988active}. Such a constraint implies that the environment will only ever be \textit{partially observed} by each acquisition. Given some prior knowledge about the parameters of the environment, however, the sensor gains the ability to infer properties of the environment without directly observing them. This process of inference on sensory states may be described as \textit{perception}, as distinct from simple measurement~\cite{van_sloun_active_2024}. We may then model this perception using the Bayesian framework, where the perceiver infers a Bayesian posterior over the parameters of the environment, with a causal model mapping those parameters to observations serving as the likelihood~\cite{kersten2004object}. The aforementioned goal of sensing may then be formalized in Bayesian terms, where $H$ is the entropy functional, \ois{$p(\cdot)$ denotes a probability density function}, $\bx$ denotes the environmental parameters to be estimated, $A$ is the set of sensing actions, and $\by$ denotes the resulting observations~\cite{rainforth2024modern}:
\begin{equation}
    \text{InfoGain}_{\bx}(A, \by) = H[p(\bx)] - H[p(\bx \mid A, \by)].
\end{equation}
In other words, the information gained by performing a sensing action $A$ is equal to the difference in uncertainty in $\bx$ before versus after observing the resulting measurements $\by$.

The perception becomes \textit{active} when the sequence of sensing actions is optimized to maximize the expected information gain, considering all the possible measurements that may result from a given sensing action \cite{rainforth2024modern}:
\begin{align}
    \label{eq:active-perception}
    A^* &=  \argmax_{A} \enspace \mathbb{E}_{p(\by \mid A)}[\text{InfoGain}_{\bx}(A, \by)] \nonumber \\
          &= \argmax_{A} I(\bx ; \by \mid A),
\end{align}
\ois{where $I(\bx ; \by \mid A)$ denotes the conditional mutual information of $\bx$ and $\by$ given $A$}. Active perception is often performed \textit{greedily}, and \textit{iteratively}, first selecting the optimal sensing action according to \eqref{eq:active-perception}, performing inference on $\bx$ given the new observations $\by$, and repeating, setting the posterior at step $t$ to the prior at step $t+1$. This process of iteratively alternating between perception and action is referred to as a \textit{perception-action loop}.

\begin{figure*}
    \centering
    \includegraphics[width=\linewidth]{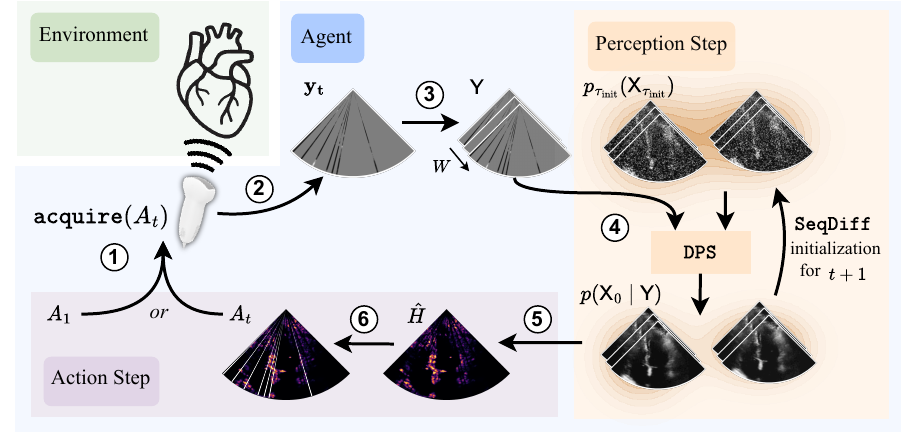}
    \vspace{-11pt}
    \caption{\ois{\circled{1} Acquire scan lines at locations specified by $A_t$ or initial locations $A_1$. \circled{2} Beamform the acquired scan lines in a zero-filled target region to produce $\by_t$. \circled{3} Place $\by_t$ in the measurement buffer $\tY$. \circled{4} Perform posterior sampling using \ac*{DPS}. \circled{5} Compute pixel-wise entropy from the \otwo{posterior} distribution. \circled{6} Select the next scan locations using \textit{K-Greedy Entropy Minimization}, and repeat.}}
    \label{fig:system-diagram}
\end{figure*}

\subsection{Posterior Sampling with Diffusion Models}\label{sec:dps}
\on{As mentioned in Section \ref{sec:active-perception}, the ability to infer Bayesian posterior distributions given partial observations is essential to perception. Given the high-dimensional nature of ultrasound video, we employ an approximate Bayesian method, performing posterior sampling with a \ac{DM}}. \acp{DM} are a powerful class of deep generative models capable of performing prior and posterior sampling of high-dimensional signals, such as images and videos~\cite{ho2020denoising, song2020score, ho2022video}. They operate by learning to reverse a corruption process wherein a sample $\bx_0 \in \mathbb{R}^N$ from the target distribution is ``diffused'' towards a Gaussian noise sample $\bep \sim \mathcal{N}(\bm{0}, \bm{I})$. This forward corruption process is modeled as follows:
\begin{equation}
\label{eq:forward-diffusion}
\bx_\tau = \alpha_\tau\bx_0 + \sigma_\tau\bep,
\end{equation}
where $\alpha_\tau$ and $\sigma_\tau$ are called the \textit{signal} and \textit{noise rates} at step $\tau$, respectively, collectively forming the \textit{diffusion schedule}. This creates a chain of samples $[\bx_0, \ldots, \bx_\tau, \ldots, \bx_{\tau_\text{max}}]$ interpolating between $\bx_0$ and $\bx_{\tau_\text{max}} = \bep$. \Acp{DM} then reverse this process iteratively, first predicting an estimate of the clean signal $\hat{\bx}_0$ from some $\bx_\tau$ using a denoising neural network, and then re-noising that estimate to a lower noise-level $\tau-1$ using the forward process \cite{song2020denoising}. This process of denoising and re-noising is repeated, refining $\hat{\bx}_0$ as $\tau \rightarrow 0$, and approaching a new random sample from the true data distribution $p(\bx)$. More formally, with an estimate of the noise $\hat{\bep}$ predicted by the denoiser, $\hat{\bx}_0$ can be computed by reversing the forward process as follows:
\begin{equation}
    \hat{\bx}_0 = \frac{1}{\alpha_\tau}(\bx_\tau - \sigma_\tau \hat{\bep}).
\end{equation}
Tweedie's formula \cite{efron2011tweedie} relates this quantity to the \textit{score function} of the marginal probability distribution over noisy samples $p_\tau(\bx_\tau)$, indicating that denoising is equivalent to taking a gradient step towards a region of higher probability density in the target distribution, in the case where $\hat{\bep}$ is produced by the minimum mean squared error denoiser:
\begin{equation}
    \hat{\bx}_0 \approx \mathbb{E}[\bx_0 \mid \bx_\tau] = \frac{1}{\alpha_\tau}(\bx_\tau + \sigma_\tau^2 \nabla_{\bx_\tau} \log p_\tau (\bx_\tau)).
\end{equation}
This notion of taking a step towards a region of higher prior probability density is referred to as the \textit{prior step}. Of particular interest in this application is Bayesian posterior sampling, wherein the model generates high-quality samples conditioned on measurements $\by \in \mathbb{R}^M$ obtained according to some known measurement model $p(\by \mid \bx)$. The \wtwo{\ac{DPS}} algorithm \cite{chung2022diffusion} solves this problem by formulating a posterior score function:
\begin{equation}
    \label{eq:posterior_score}
    \underbrace{\nabla_{\bx_\tau} \log p_\tau(\bx_\tau | \by)}_{\text{posterior}} = \underbrace{\nabla_{\bx_\tau}\log p_\tau(\bx_\otwo{\tau})}_{\text{prior}} + \underbrace{\nabla_{\bx_\tau} \log p_\tau (\by | \bx_\tau)}_{\text{likelihood}}.
\end{equation}
The likelihood term in \eqref{eq:posterior_score} is derived from a known measurement model, typically with some additive noise, e.g. $p(\by \mid \bx) = \mathcal{N}(\by; \mathcal{A}(\bx), \sigma_{\bn}^2\bm{I})$, where $\mathcal{A}$ is some measurement operator. \ac{DPS} then approximates the likelihood score at step $\tau$ using the Tweedie estimate $\hat{\bx}_0$ computed during the prior step. With Gaussian measurement noise, this becomes:
\begin{equation}
    \label{eq:dps}
    \nabla_{\bx_\tau} \log p_\tau(\by \mid \bx_\tau) \simeq -\frac{1}{\sigma_{\bn}^2}\nabla_{\bx_\tau}||\by - \mathcal{A}(\hat{\bx}_0)||_2^2.
\end{equation}
Adding the gradient in equation \eqref{eq:dps} to $\bx_\tau$ constitutes the \textit{likelihood step}. \ac{DPS} alternates between prior and likelihood steps during inference, leading to samples that accord with the measurements while remaining plausible under the prior.

\section{Method}
\label{sec:method}
In this section, we present CASL in terms of its two primary components: (i) \textit{perception}, in which a posterior distribution over the possible states of the tissue is inferred from a partial observation, and (ii) \textit{action}, in which this perceived distribution is used to select the next transmit lines. \wn{An overview of the method is shown in \autoref{fig:system-diagram}.}

\subsection{Perception}
\otwo{Perception, as defined in Section \ref{sec:active-perception}, can be formalized as Bayesian posterior inference. In our case, this amounts to inferring} the tissue state $\bx_t$ at time $t$ given the history of observations and actions until that point, i.e., the distribution $p(\bx_t \mid \bh_t)$, where $\bh_t$ indicates the observation history at time $t$, consisting of the actions $A_{1:t}$ taken so far, and their resulting observations $\by_{1:t}$. \ois{Concretely, the tissue state $\bx_t \in \mathbb{R}^N$ is represented by a fully-sampled image, and the measurements $\by_t \in \mathbb{R}^N$ are partially-sampled images containing zeros at unmeasured tissue locations.} \otwo{Functionally, the perception step in CASL produces two important quantities. Firstly, a point estimate (e.g. a posterior sample or the posterior mean), which can serve as the \textit{reconstruction} image for a given frame, and secondly, a pixel-wise uncertainty map (derived from multiple samples) which is used to drive information-maximizing action selection, discussed in Section~\ref{sec:action}. Both of these quantities can be estimated from a set of posterior samples, motivating our use of the \ac{DPS} algorithm, introduced in Section~\ref{sec:dps}.} Given that ultrasound video exhibits strong temporal dependencies between frames, it is important to model the conditional relationship between $\bx_t$ and past measurements $\by_{1:t}$. To model such dependencies, we fit the diffusion model on sequences of $W$ consecutive frames $\tX =[\bx_{t-W+1}, \dots, \bx_t]$ sampled at random from the training set, learning a prior over tensors $\tX \in \mathbb{R}^{N \times W}$. In other words, the model has a temporal context window of size $W$. This amounts to a prior model with a $W$-order Markov assumption on ultrasound video, where $W$ can be chosen to balance the benefits in predictive ability with the cost of increasing training data sparsity and inference compute as $W$ increases. For the models presented in this work, we use $W=3$. \ois{The denoiser $\bep_\theta$ is implemented using a U-Net architecture \cite{ronneberger2015u} \wtwo{of 2M parameters}, with further details available on the GitHub repository.} 

During inference, at each time step $t$ we generate a set of \ois{$N_p$} tensors \ois{$\{\tX_0^{(i)}\}_{i=1}^{N_p}$} in parallel. The final image \ois{$\tX_0^{(i)}[W]$} in each generated tensor represents one possible state of $\bx_t$. These images, dubbed \textit{particles}, can then be used to approximate the posterior distribution $p(\bx_t \mid \bh_{t})$. Throughout the paper, we refer to this set of particles $\{\bx_t^{(i)}\}_{i=1}^{N_p}$ as the agent's \otwo{posterior distribution} at time $t$, with differences across particles indicating uncertainty in the state of $\bx$. Throughout our experiments, we use \wes{$N_p=2$, leading to minimal computational overhead.}

We must then specify a likelihood function to guide generation with \ac{DPS}. We start by stacking our acquired scan line measurements in a measurement buffer $\tY=[\by_{t-W+1}, \dots, \by_t]$. Then, we define a measurement model simulating focused line-scanning. This model assumes that for each focused transmit, a set of pixels extending along the focus line is beamformed, and that a frame is created by concatenating a string of such lines. The measurement model is thus a masking operation, wherein the full frame is mapped to a set of measurements by revealing only those that were acquired. In particular, $\tA \in \mathbb{R}^{N\times W}$ is a measurement mask extending across the context window containing ones at the pixel locations measured by the acquired scan lines, and zeros elsewhere. \on{Since this measurement model is deterministic, its likelihood is a Dirac delta distribution, i.e., $p(\tY | \tX, \tA)~=~\delta(\tY - \tA \odot \tX)$\ois{, where $\odot$ denotes an element-wise product}. To ensure smooth gradients for \ac{DPS}, however, we instead use a Gaussian distribution, which is a continuous relaxation of the Dirac delta. This yields the following likelihood, where the variance $\sigma_{\bn}^2=\gamma^{-1}$ is a hyperparameter:
\begin{equation}
    \label{eq:measurement-model}
    p(\tY \mid \tX, \tA) = \mathcal{N}(\tY ; \tA \odot \tX, \sigma_{\bn}^2 \otwo{\bm{I}}).
\end{equation}
Computing the score of this likelihood function produces the following guidance step in \ac{DPS} for diffusion step $\tau$:
\begin{equation}
      \nabla_{\tX_\tau} \log p_\tau(\tY \mid \tX_\tau) \simeq - \gamma\nabla_{\tX_{\tau}}||\tY - \tA \odot \hat{\tX}_0||_2^2 .
\end{equation}}
In the case where the beamforming grid is specified in the polar domain, we fit the diffusion model on polar domain data, such that the model remains the same on polar and Cartesian grids, in each case simply revealing or masking vertical lines of pixels. \on{In order to accelerate inference and create a temporally consistent video, we employ SeqDiff~\cite{stevens2025sequential} initialization}. \on{Finally, we return for each frame a single \textit{reconstruction} image, $\tilde{\bx}_t$, which is chosen to be the first particle $\tilde{\bx}_t := \bx_t^{(1)}$ of the \otwo{posterior} distribution.} \ois{The first particle is chosen \otwo{as a random sample} from the \otwo{posterior} distribution to ensure that the reconstruction image is on-manifold~\cite{chung2022diffusion}, enhancing perceptual quality. Alternate reconstruction functions may, however, be used depending on the goal, for example, to minimize distortion we can approximate the posterior mean by averaging posterior samples \cite{blau2018perception}. Finally, we \textit{hard-project} the measurements onto the reconstruction, setting the values at measured pixels to be identical to those in the measurement}. \wes{Throughout this work, we compute image-quality metrics on the images in the polar domain, before scan conversion.}

\subsection{Action}
\label{sec:action}
The action step aims to choose a set of actions to take at time~$t~+~1$ given the \otwo{posterior} distribution at time $t$. \otwo{An action in this case corresponds to firing a focused transmit beam at a particular steering angle}. \otwo{The action space is therefore represented by} a discretized set of possible focused scan locations $\{A^\ell\ \mid \ell = 1, 2, \dots, L\}$, where there are $L$ total locations. \otwo{We denote by} $A^\ell$ the set of indices of the pixels that are measured by that action, facilitating the creation of a corresponding measurement mask~$\mathcal{M}(A^\ell)$, \on{where $\mathcal{M}$ creates a matrix containing} ones at the indices specified by $A^\ell$ and zeros elsewhere. \otwo{For brevity, we use the term \textit{action} to refer to the pixel indices $A^\ell$ corresponding to that action, and we use a time subscript $A_t^\ell$ to denote a specific action to be taken at time $t$.} The actions should be chosen to maximize information gain with respect to the tissue state, following the objective described in Section~\ref{sec:active-perception}. Starting with the expected information gain objective provided in~\eqref{eq:active-perception}, and following van Sloun~\cite{van_sloun_active_2024}, we derive our action selection policy, substituting in the likelihood function specified in~\eqref{eq:measurement-model}. \ois{Because the conditional mutual information is symmetric, we can compute the expected reduction in uncertainty about $\bx_t$ in terms of uncertainty about the observations under future actions, avoiding the need for an additional perception step on hypothetical observations. Then, for computational efficiency, rather than optimizing directly for the future action $A^\ell_{t+1}$, which would require simulating a future \otwo{posterior} distribution, we make use of the present \otwo{posterior} distribution, and instead optimize for the hypothetical action $A^{\ell'}_t$ that would result in the most informative additional observation $\by'_t$ given the history $\bh_t$, leading to the following measure of information gain:}
\begin{align}
\label{eq:action-selection}
I(\by'_t ; \bx_t \mid \bh_t) \nonumber
&=  H(\by'_t \mid A_t^{\ell'}, \bh_{t}) - H(\by'_t \mid \bx_t, A_t^{\ell'}, \bh_{t}) \nonumber \\
&= H(\by'_t \mid A_t^{\ell'}, \bh_{t}) - H(\bn).
\end{align}
The second entropy term $H(\by'_t \mid \bx_t, A_t^{\ell'}, \bh_t)$ is the entropy of our likelihood function, whose only source of uncertainty is the additive noise $\bn$. $H(\bn)$ then drops out when we take the argmax with respect to the action $A_t^{\ell'}$, yielding the following objective, selecting the most informative location $\ell'$:
\begin{equation}
    \argmax_{\ell'} I(\by'_t ; \bx_t \mid A_t^{\ell'}, \bh_t) = \argmax_{\ell'} H(\by'_t \mid A_t^{\ell'}, \bh_t).
\end{equation}
\on{The remaining entropy values for each line measurement $H(\by'_t~|~A_t^{\ell'},~\bh_t)$ can be decomposed into sums of pixel-wise entropies by modeling the pixels as independent variables. Given that pixels masked by $A_t^{\ell'}$ have zero entropy, the measurement entropy can be computed as a function of pixel entropies in $\bx_t$, where $\bx_t[\sfi]$ denotes the $i^{th}$ pixel of $\bx_t$:}
\begin{equation}
    H(\by'_t~|~A_t^{\ell'},~\bh_t) = \sum_{\sfi \in A_t^{\ell'}} H(\bx_t[\sfi] \mid A_t^{\ell'}, \bh_t).
\end{equation}
In practice, we first compute a pixel-wise entropy map in the image domain $\bx_t$, $\hat{H} = [\hat{H}[1], \dots, \hat{H}[\sfi], \dots, \hat{H}[N]]^\top$, where $\hat{H}[\sfi] \approx H(\bx_t[\sfi] \mid A_t^{\ell'}, \bh_{t})$. Given $\hat{H}$, we can sum the pixels corresponding to each action $A_t^{\ell'}$ in order to get the line-wise measurement entropies $\hat{H}^\ell$, choosing the maximum entropy line as the next action. Using the variational entropy approximation proposed by Hershey \textit{et al.} \cite{hershey2007approximating}, the pixel-wise entropy map $\hat{H}$ can be computed by taking the element-wise squared error between each pair of particles in the \otwo{posterior} distribution $\{\bx_t^{(i)}\}_{i=1}^{N_p}$, as follows\otwo{, where the $\exp$ and $\log$ functions are applied element-wise}:
\begin{equation}
    \hat{H} =  -\sum_{i=1}^{N_p} \frac{1}{N_p} \log \sum_{j=1}^{N_p} \frac{1}{N_p} \exp \left[- \frac{(\bx_t^{(i)}-\bx_t^{(j)})^2}{2\sigma_{\bx}^2} \right].
\end{equation}
Intuitively, this entropy map will have high values in regions where the images in the \otwo{posterior} distribution \textit{disagree} with one another, indicating uncertainty. Selecting the maximum entropy line $\ell'$ from this entropy map then amounts to:
\begin{equation}
    \argmax_{\ell'} H(\by'_t \mid A_t^{\ell'}, \bh_t) \approx \argmax_{\ell'} \sum_{\sfi\in A_t^{\ell'}}\hat{H}[\sfi].
\end{equation}

We could proceed with the above as our policy, selecting one line at a time, performing the perception step for the resulting measurement, and repeating. However, the perception step requires executing some reverse diffusion steps, decreasing the frame rate. In order to prevent this, we propose an approximate algorithm, called \textbf{K-Greedy Entropy Minimization}, where we batch the measurements that will be used to generate each frame and perform perception on that batch. This algorithm approximates the decrease in entropy that would result from conditioning on a given measurement using a \ac{RBF} around the measurement location. This effectively assumes that measuring a line~$\ell$ will provide information about nearby lines, decreasing exponentially with distance. The algorithm proceeds by selecting the maximum entropy line, reweighting the entropies of the neighboring lines according to the \ac{RBF}, and repeating, for $K$ total lines. For a formal presentation of this algorithm, see the \textit{action} step in Algorithm~\ref{alg:perception-action}. \otwo{In the algorithm, we use the notation $A_t$ to gather the set of pixel indices from all selected lines.}

\algnewcommand{\RequireOptional}{\item[\textbf{Optional:}]}
\algrenewcommand\algorithmiccomment[1]{\hfill\textcolor{gray!60}{//~#1}}
\makeatletter %
\newcommand{\CommentLine}[1]{%
  \Statex \hspace{\ALG@thistlm}\textcolor{gray!60}{//~#1}%
}
\makeatother %

\definecolor{perceptioncol}{HTML}{FFE6CC}
\definecolor{actioncol}{HTML}{E1D5E7}

\makeatletter
\newcommand{\CommentBlock}[2]{%
  \Statex\vspace{4pt}\hspace{\ALG@thistlm}%
  \tcbox[
    colback=#1,                %
    coltext=black,             %
    arc=3pt,
    boxrule=0pt,
    boxsep=2pt,
    left=-1pt, right=2pt,
    top=1pt, bottom=1pt,
    nobeforeafter
  ]{~#2}%
}
\makeatother

\begin{algorithm}
\caption{CASL Perception-Action Loop}
\label{alg:perception-action}
\begin{spacing}{1.2}
\begin{algorithmic}[1]  %
\Require \texttt{SeqDiff} initial diffusion step $\tau_\texttt{SeqDiff}$; \enspace total diffusion steps $\tau_\text{max}$; \enspace number of focused transmit locations~$L$; \enspace number of particles~$N_p$; \enspace number of focused transmits per frame~$K$; \enspace initial transmit indices~$A_1$; \enspace diffusion schedule $\{\alpha_{\tau},\sigma_{\tau}\}_{\tau=0}^{\tau_\text{\otwo{max}}}$; \enspace guidance weight~$\gamma$; \enspace posterior variance~$\sigma^2_{\bx}$; \enspace temporal window size~$W$; \enspace \ois{$\forall t < 1: \bx_t \leftarrow \mathbf{0}, \by_t \leftarrow \mathbf{0}$}.
\Ensure Sequence $\{ \tilde{\bx}_t \}_{t=1}^T$ of reconstructed frames.
\For{$t \in [1, ..., T]$}
    \State $\by_t \gets \texttt{acquire}(A_t)$ \Comment{Acquire measurements} 
    \State $\tY \gets [\by_{t-W+1}, \dots, \by_t]$ \Comment{Measurement buffer}
    \State $\tA \gets [\mathcal{M}(A_{t-W+1}), \dots, \mathcal{M}(A_t)]$ \Comment{Mask buffer}
    \If{$t = 1$}
        \State $\tau_\text{init} \gets \tau_\text{max}$
    \Else
        \State $\tau_\text{init} \gets \tau_\texttt{SeqDiff}$ 
    \EndIf
    \CommentBlock{perceptioncol}{Perception Step}
    \For{each $i \in \{1, ..., N_p\}$ in parallel}
        \State $\tX \gets [ \bx^{(i)}_{t-W}, \dots, \bx_{t-1}^{(i)}]$ \Comment{$\tX$ shorthand for $\tX_t^{(i)}$}
        \State $\bep \sim \mathcal{N}(\textbf{0}, \bm{I})$ \Comment{Initial noise}
        \State $\tX_{\tau_{\text{init}}} \gets \alpha_{\tau_{\text{init}}} \tX + \sigma_{\tau_{\text{init}}}\bep$ \Comment{Initial samples}
        \For{$\tau \in [\tau_{\text{init}}, ..., 0)$}
            \State $\hat{\bep} \gets \bep_\theta(\tX_{\tau}, \sigma^2_\tau)$ \Comment{Predict noise}
            \State $\hat{\tX}_0 \gets (\tX_\tau - \sigma_\tau \hat{\bep}) / \alpha_\tau$ \Comment{Tweedie Estimate}
            \State $\tX'_{\tau-1} \gets \alpha_{\tau-1}\hat{\tX}_0 + \sigma_{\tau-1}\hat{\bep}$ \Comment{Prior step}
            \State $\tX_{\tau-1} \gets \tX'_{\tau-1} - \gamma\nabla_{\tX_{\tau}}||\tY - \tA \odot \hat{\tX}_0||_2^2$
        \EndFor
        \State $\bx_{t}^{(i)} \gets \tX_0[W]$ \Comment{Posterior distribution $t$}
    \EndFor
    \State $\tilde{\bx}_t \gets \bx_t^{(1)}$ \Comment{Choose first as reconstruction}

    \CommentBlock{actioncol}{Action Step}
    \State $A_{t+1} \gets \emptyset$ \Comment{Initialize action set for next transmit}
    \State $\hat{H}\gets  -\sum_i \frac{1}{N_p} \log \sum_j \frac{1}{N_p} \exp \left[- \frac{(\bx_t^{(i)}-\bx_t^{(j)})^2}{2\sigma_{\bx}^2} \right] $ %
    \State $\forall \ell: \hat{H}^\ell \gets \sum_{\sfi \in A^\ell}\hat{H}[\sfi]$ \Comment{Line-wise entropy}
    \For{$k \in [1, \dots, K]$}
        \State $\ell^* \gets \underset{\ell}{\argmax}\text{ }\hat{H}^\ell$ \Comment{Select max entropy action}
        \State $A_{t+1} \gets A_{t+1} \cup A^{\ell^*}$ \Comment{Gather selected actions}
        \State $\otwo{\forall \ell:} \hat{H}^\ell \gets \hat{H}^\ell * \left(1-\exp\left(-\frac{(\ell - \ell^*)^2}{2} \right)\right)$
    \EndFor
\EndFor
\Statex
\Return $\{ \tilde{\bx}_t \}_{t=1}^T$
\end{algorithmic}
\end{spacing}
\end{algorithm}

\begin{figure*}
    \centering
    \includegraphics[width=\linewidth]{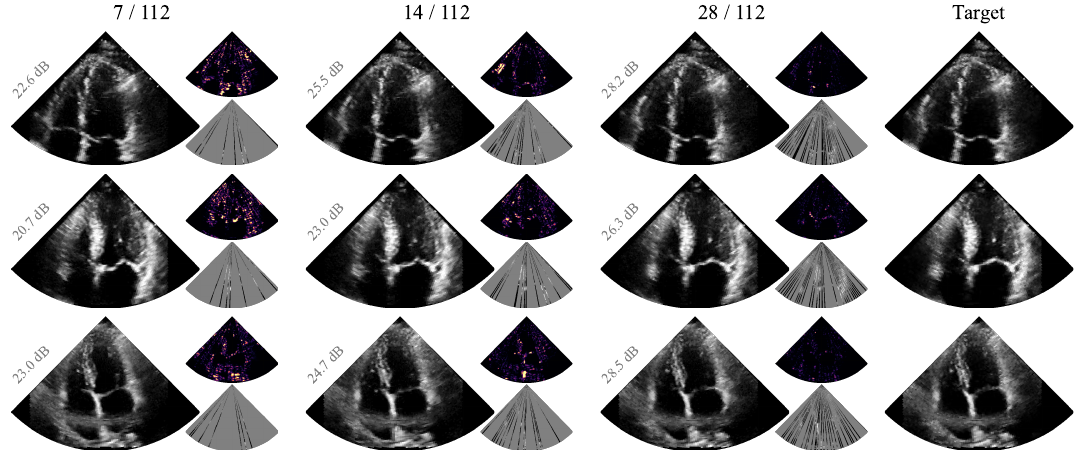}
    \vspace{-9pt}
    \caption{Qualitative results on the EchoNet-Dynamic dataset. The figure shows the reconstructions \wes{for 7, 14, and 28 lines compared to the target, for three different subjects in the test set. Additionally, it shows the reconstruction quality, in terms of \acs*{PSNR}, the acquired lines, and the posterior entropy, which drives action selection.}}
    \label{fig:qualitative_echonet}
\end{figure*}

\section{Experiments}
\label{sec:experiments}
A comprehensive evaluation of \ois{CASL}'s performance is provided through a series of experiments. 
\wn{First, we test our method on the EchoNet-Dynamic dataset, which is an image dataset from which we simulate subsampling transmits} \on{using a masking measurement model. }\wn{Next, we use an in-house dataset where we can directly subsample the transmit events in the channel data, and beamform those transmits to independent lines. Lastly, we show that our method can also be applied to 3D echocardiography, where we subsample elevation planes. \wtwo{In all experiments we retrospectively subsample all lines to simulate acquiring only the selected lines.} \ois{CASL} will be compared to equispaced and random subsampling, using the same diffusion model \ois{for perception}. The equispaced subsampler ``rolls'' the selected lines from left to right, such that over time the full imaging area is measured. Random sampling means that the selected lines were sampled from a uniform distribution. We implement \ois{CASL} using \texttt{zea}, the cognitive ultrasound toolbox \cite{zea2025}.}

\subsection{EchoNet-Dynamic} \label{sec:echonet-dynamic}
Here we train a diffusion model on the EchoNet-Dynamic dataset \cite{echonet_dynamic}. The EchoNet-Dynamic dataset consists of over 10k echocardiograms. As we do not have access to how the data was beamformed or the channel data, we opted to simulate scan lines as a column of pixels of the 112$\times$112 images. To that \otwo{end}, we have converted the dataset from scan-converted images back to the polar domain. \wn{In the process, we excluded 2,044 samples because their scan-converted images were generated using a different method or parameters, which prevented consistent conversion to the polar format used for the rest of the dataset.} The rest of the data we have randomly split on the patient level into 6985 train sequences, 500 validation sequences, and 500 test sequences. While we used the full sequences to train our model, we use 100 frames per patient for the metrics to ensure every patient gets weighted equally in the metrics.

\subsubsection{Reconstruction quality}
The qualitative results are shown in \autoref{fig:qualitative_echonet}. Here, the 20th frame is used for three random patients in the test data. \wes{We show reconstructions for three subsampling rates. Additionally,} we show the acquired lines, the entropy of the posterior samples, and the fully observed target images. The reconstructions are visually very similar to the targets, while \wes{in the extreme case,} using only 7 out of 112 scan lines.

\begin{figure}
    \centering
    \includegraphics[width=\linewidth]{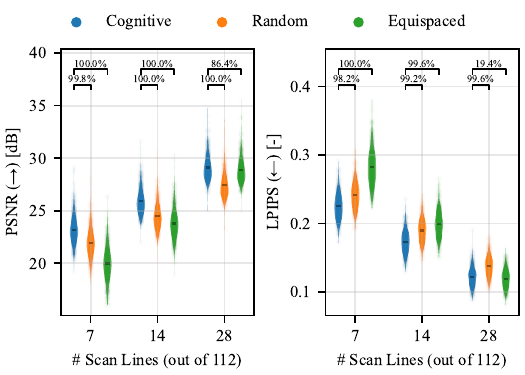}
    \vspace{-9pt}
    \caption{Reconstruction performance for EchoNet-Dynamic in terms of \acs*{PSNR} and \acs*{LPIPS} \wes{\cite{zhang_unreasonable_2018}} as a function of the number of scanned lines for various action selection policies. The figure shows a distribution over the data samples and includes the mean as a gray line. \wtwo{The brackets show the win-rate over the baselines. Cognitive is found to be significantly better ($p<10^{-53}$) than the other baselines for all subsampling rates and metrics using the Wilcoxon signed-rank test.}}
    \label{fig:echonet_psnr}
\end{figure}

\autoref{fig:echonet_psnr} shows the reconstruction quality in terms of \ac{PSNR} and \ac{LPIPS} \wes{\cite{zhang_unreasonable_2018}} as distribution\on{s} over all the patients in the test dataset. \wtwo{The brackets in the figure indicate the win-rate of our method over the baselines.} It can be seen that cognitive subsampling outperforms the other subsampling strategies\wn{, especially for lower subsampling rates.} For 7 out of 112 lines, which is just over 6\% of the image, \wtwo{CASL} still achieves a \ac{PSNR} of 23.2 on average, which consists of a 5.7\% improvement over random sampling and an impressive 16.2\% improvement over equispaced sampling. \wtwo{The strong significance is explained by the result that CASL outperforms baselines for both metrics in virtually every cine loop.}

\subsubsection{Hyperparameters}
\begin{figure}
    \centering
    \includegraphics[width=\linewidth]{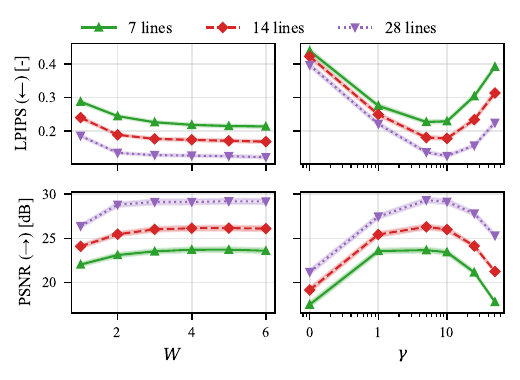}
    \vspace{-9pt}
    \caption{\ois{Reconstruction quality on EchoNet-Dynamic validation samples as a function of hyperparameter values $W \in [1, 2, 3, 4, 5, 6]$ and $\gamma \in [0.1, 1, 5, 10, 25, 50]$, while keeping the remaining parameters fixed. The mean result across 20 patients from the validation set with 100 frames each is plotted, with shaded regions indicating the standard error of the mean. For $W$, performance plateaus at some point, depending on the number of scan lines, while $\gamma$ exhibits a convex curve for both metrics with optima at $\gamma \approx 10$.}}
    \label{fig:sweeps}
\end{figure}
\begin{figure}
    \centering
    \includegraphics[width=\linewidth]{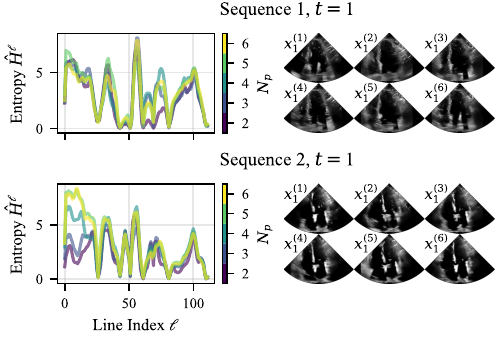}
    \vspace{-15pt}
    \caption{\ois{Linewise entropy estimates generated using different numbers of particles $N_p$ for the same input measurements on the first frame of two different sequences from the EchoNet-Dynamic validation set, where uncertainty will be highest. It is clear that while some differences exist between the estimates, the peaks lie in similar locations, leading to similar actions.}}
    \label{fig:linewise-entropy}
\end{figure}

\ois{Algorithm \ref{alg:perception-action} relies on a number of hyperparameters, most notably the temporal window size $W$, guidance weight $\gamma$, and number of particles $N_p$. In order to measure the influence of these parameters on reconstruction quality, a hyperparameter sweep was performed for each on a set of 20 patients from the EchoNet-Dynamic validation set with 100 frames each. In each case, all parameters aside from the one being swept over are fixed to the optimal values. Both $W$ and $\gamma$ were found to exert some influence on reconstruction quality, as illustrated in \autoref{fig:sweeps}. For $W$, we find that the quality plateaus at $W=3$, thereby making $W=3$ a sensible choice, since the required compute increases with $W$. We do however observe that the plateau appears later as the number of scan lines decreases, indicating that models with longer context windows may improve performance under more extreme undersampling. $\gamma$ shows convex curves for both \ac{PSNR} and \ac{LPIPS} with optima around $\gamma=10$. Interestingly, $N_p$ was found to have a negligible effect on reconstruction performance, with $N_p=2$ performing almost identically to $N_p \in \otwo{\{}3, 4, 5, 6\otwo{\}}$. One explanation for this is the coarse nature of the action space, which sums the entropy along an entire scan line, meaning that subtle differences between particles along the vertical axis do not affect the overall linewise entropy estimates. This effect is illustrated by Fig. \ref{fig:linewise-entropy}, which plots the linewise entropy estimates for the same frame using increasing values for $N_p$. It is clear in the figure that the entropy estimates are similar for increasing $N_p$, leading to similar actions. The result is that using $N_p=2$ is sufficient to drive uncertainty-minimizing scan line selection without requiring excessive computational resources. \otwo{While these entropy estimates lead to similar actions, it is worth commenting on \textit{why} they differ for different $N_p$. Since the particles are randomly drawn from the posterior distribution, it can be that the set of particles agree with one another in a certain region \textit{by coincidence}, resulting in an underestimation of the entropy in that region. Similarly, particles could coincidentally disagree in a region due to an unlikely mode of the distribution being over-sampled.}}

\begin{figure}
    \centering
    \includegraphics[width=\linewidth]{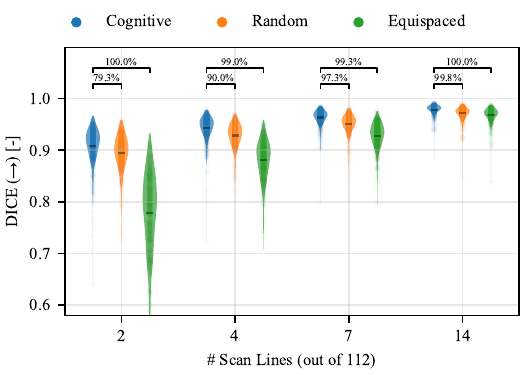}
    \vspace{-9pt}
    \caption{Segmentation performance in terms of \acs*{DICE} of EchoNet-Dynamic on subsampled images for various action selection policies. The figure shows a distribution over the data samples and includes the mean as a gray line. \wtwo{The brackets show the win-rate over the baselines. Cognitive is found to be significantly better ($p<10^{-29}$) than the other baselines for all subsampling rates and metrics using the Wilcoxon signed-rank test.}}
    \label{fig:segmentation_dice}
\end{figure}

\begin{figure*}
    \centering
    \includegraphics[width=\linewidth]{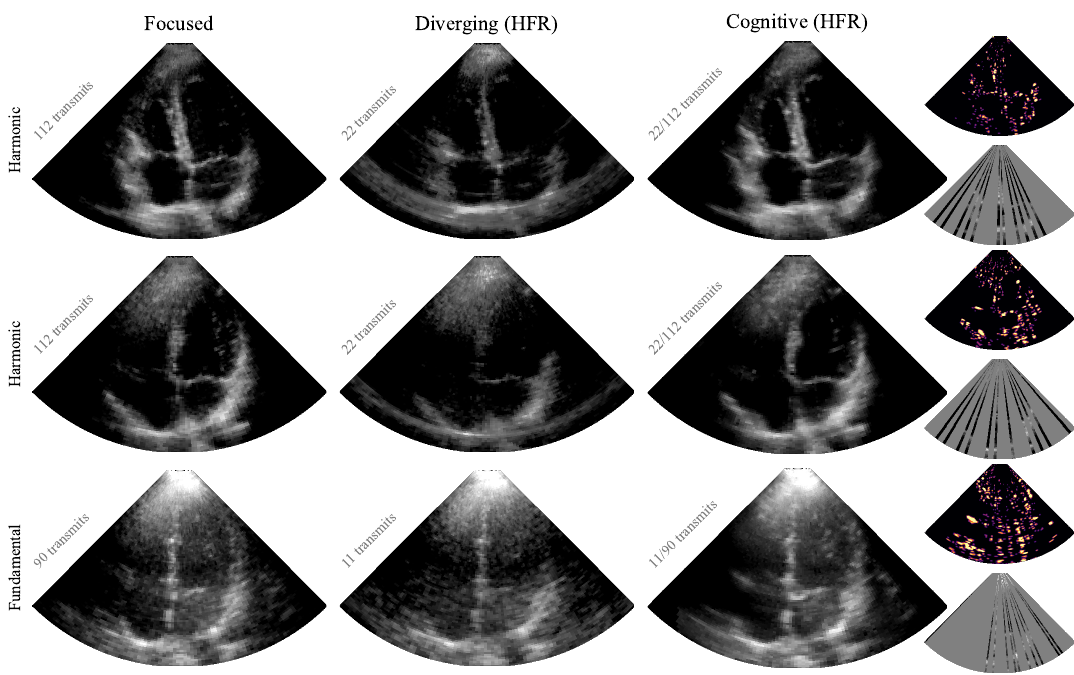}
    \vspace{-9pt}
    \caption{Qualitative results on the in-house echocardiograms. \wes{The top rows show frames from an echocardiogram using harmonic imaging; the third shows another subject using fundamental imaging. During the scan, both focused and diverging waves were acquired.}
    \wes{Each row shows the fully observed focused acquisition, \acf*{HFR} acquisition using diverging waves, and \acs*{HFR} using focused cognitive transmits.}
    On the right, the acquisitions\wes{, entropy and} reconstructions are shown for \wes{the subsampled} focused transmits. All images were 112$\times$112 pixels before scan conversion \wes{and histogram matched to the fully observed focused data.}}
    \label{fig:inhouse}
\end{figure*}

\subsubsection{Left ventricle segmentation}
A common parameter extracted from an echocardiogram is the ejection fraction, which measures the amount of blood pumped out of the heart's left ventricle with each heartbeat. The EchoNet-Dynamic model~\cite{echonet_dynamic} can segment the left ventricle with high accuracy. In this experiment, we will evaluate how the subsampled reconstructions affect the ability to segment the left ventricle. We will use the \wes{\ac{DICE}} to compare the segmentations of the subsampled images and the fully observed images. We exclude failure cases from the fully observed image sequences in which the segmentation model generates multiple disconnected components in at least five consecutive frames. \autoref{fig:segmentation_dice} shows that \ois{CASL} consistently produces the best left ventricle segmentations compared to equispaced and random subsampling. The performance for 2 out of 112 \wtwo{lines} still reaches a \ac{DICE} of 0.91 on average.

\begin{table}
\caption{Inference speed optimizations computed on the \wes{NVIDIA L40S} for 112$\times$112 pixels \wes{and $N_p=2$}.}
\renewcommand{\arraystretch}{1.1}
\label{tab:inference_speed}
\begin{tabular}{p{4.4cm}|cc}
\toprule
 & \multicolumn{2}{c}{\textbf{Frame}} \\
\textbf{Optimization} & \textbf{Time [ms]} & \textbf{Freq. [Hz]} \\ \midrule
\west{Base} (\textit{500 steps}) & \west{858.7} & \west{1.16} \\
+ SeqDiff \cite{stevens2025sequential} (\textit{25 steps}) & \west{76.0} & \west{13.16} \\
+ Just-in-time compilation & \west{36.0} & \west{27.81} \\
+ Mixed precision (\textit{float16}) & \west{17.3} & \west{\textbf{57.65}} \\ 
\midrule
\west{Physical \textit{subsampled}} acquisition (28 \west{transmits}) & 5.46 & 183.2 \\
\west{Physical \textit{full} acquisition (112 transmits)} & \west{21.82} & \west{45.83} \\
\bottomrule
\end{tabular}
\end{table}

\subsubsection{Inference speed} \label{sec:inference_speed}
As mentioned before, we employ SeqDiff~\cite{stevens2025sequential}, which not only improves temporal consistency of posterior samples, but it also massively reduces the required number of \ois{neural} function evaluations \ois{to generate} sequential signals.
To improve inference speed further, we applied a group of optimizations as shown in \autoref{tab:inference_speed}. First, we chose 25 SeqDiff steps as a good balance between reconstruction quality and inference speed. Then we applied just-in-time compilation \wes{to the function encapsulating both the ``Perception Step'' and ``Action Step''} using the JAX library \cite{jax2018github}. 
Finally, the diffusion model, trained in 32-bit floating point precision, can be run in mixed precision using 16 bits. \wes{All these optimizations combined yield a frame rate of 58 Hz on the Nvidia L40S. \wtwo{We measured GPU memory usage to be approximately 720 MB. The inference speed of the reconstruction model is independent of the number of transmits.}
The physical time it takes to acquire 112 transmits is 21.82 ms, meaning the maximal achievable frame rate is 46 Hz. Given that our model can run faster, this would enable increasing the frame rate to 58 Hz through subsampling 89/112 focused transmits. We have also listed the frame rate for 28 transmits, indicating that improvements in inference speed could further increase frame rates.
}

\subsection{In-house echocardiograms}
The in-house dataset \wes{(\autoref{fig:inhouse}) was recorded on a Verasonics Vantage 256 with an S5-1 Philips transducer with a center frequency of 3.125~MHz. The scans were performed by an experienced sonographer. The dataset consists of six} subjects, who provided informed consent at the time of data collection, and the study was approved by the local Institutional Review Board. \wes{We included two types of acquisitions: fundamental mode imaging was used for three subjects, and harmonic imaging with pulse inversion was used for the other three subjects.} The \wes{fundamental} acquisition consists of 90 focused transmits, which were interleaved with 11 diverging transmits for comparison\wes{, at a \ac{PRF} of 4.32~kHz, and the transmit frequency is equal to the probe's center frequency. The harmonic acquisition consists of 56 $\times$ 2 focused transmits, one for each polarity, interleaved with 11 $\times$ 2 diverging transmits, at a \ac{PRF} of 5.07~kHz. The transmit frequency is 1.9531~MHz, and the demodulation frequency is double the transmit frequency, to capture the harmonic signal. The recorded sequences consist of \wes{70 to} 100 frames in the apical four-chamber view. The dynamic range is matched to the EchoNet-Dynamic dataset using percentile-based clipping.} \ois{Subsampling is implemented by taking a subset of} transmit events from the channel data and independently beamforming only those transmit events to \wes{lines. The width of the beamformed lines is 1 or 2 pixels for the fundamental and harmonic data, respectively. Both modalities are beamformed to \wtwo{2} pixels per wavelength axially. Subsequently, \wes{the lines are downsampled} to \wes{112 pixels, using linear interpolation and a triangular filter for anti-aliasing}, to match the training dataset}, giving us~$\by_t$. The pretrained \wes{(EchoNet-Dynamic)} prior \ois{was} used to generate reconstructions~$\tilde{\bx}_t$.

\subsubsection{Contrast}

\wn{To demonstrate the effectiveness of our method, we compute the \ac{gCNR} metric \wes{\cite{rodriguez-molares_generalized_2020}} between the ventricle and the myocardium as well as between the ventricle and the valve. \wes{These \acl{ROIs} are manually annotated, using the fully observed image (all focused transmits), by one of the authors. The same annotations are used for all the reconstruction methods.} The \ac{gCNR} is calculated relative to the fully sampled focused acquisition, which allows us to compare \ois{CASL} to diverging waves for the same number of transmits.}

\begin{figure}
    \centering
    \includegraphics[width=\linewidth]{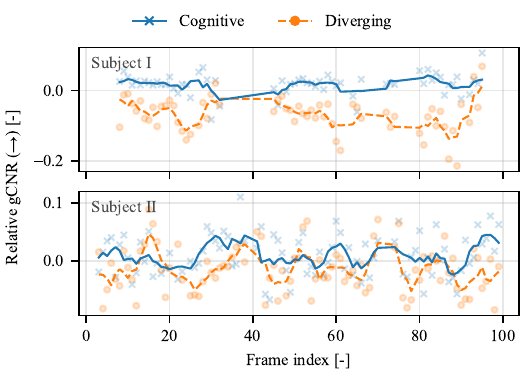}
    \vspace{-9pt}
    \caption{\Acf*{gCNR} for two subjects over time relative to a focused acquisition of 90 transmits. The \acs*{gCNR} was measured between the \textbf{valve} and the \textbf{ventricle}. Both \ois{cognitive} and diverging use 11 transmits.}
    \label{fig:gcnr_valve_over_time}
\end{figure}

\begin{figure}
    \centering
    \includegraphics[width=\linewidth]{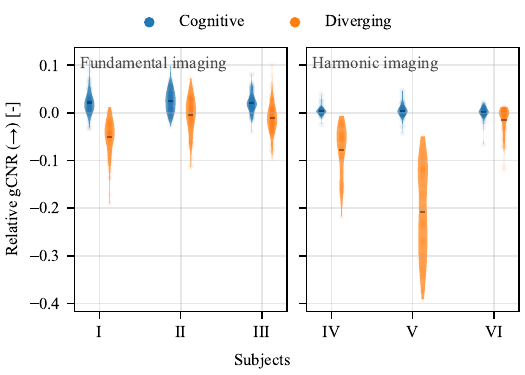}
    \vspace{-9pt}
    \caption{\Acf*{gCNR} for \wes{six} subjects relative to a focused acquisition. The \acs*{gCNR} was measured between the \textbf{myocardium} and the \textbf{ventricle}. Both \ois{cognitive} and diverging use \wes{the same number of transmits, 11 for fundamental and 22 for harmonic}. The figure shows a distribution over the frames and includes the mean as a gray line. \wtwo{Cognitive is found to be significantly better ($p<0.05$) than diverging using the Wilcoxon signed-rank test.}}
    \label{fig:gcnr_violin}
\end{figure}

\wn{\autoref{fig:gcnr_valve_over_time} shows the \ac{gCNR} over time between the valve and the ventricle for two subjects. It can be seen that \ois{CASL} almost always outperforms diverging waves. \ois{CASL} generally has slightly higher \ac{gCNR} compared to focused transmits, while for diverging waves it is slightly lower. In \autoref{fig:gcnr_violin} we show the distribution of \ac{gCNR} over the frames between the myocardium and ventricle for \wes{six} subjects. This highlights again that \ois{CASL} outperforms diverging waves for all subjects, and shows fewer outliers.}

\begin{figure}
    \centering
    \includegraphics[width=\linewidth]{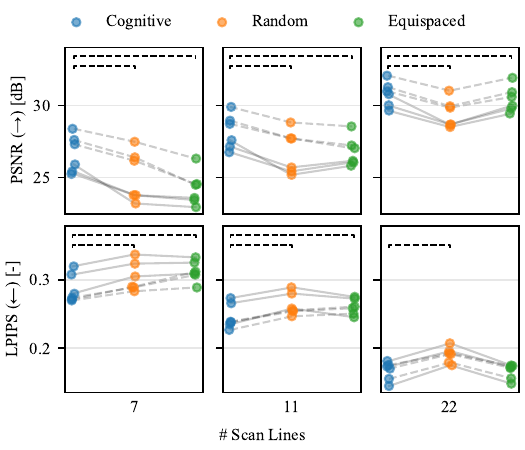}
    \vspace{-9pt}
    \caption{\wes{Reconstruction performance for six subjects in terms of \ac*{PSNR} \wtwo{and \ac*{LPIPS} \cite{zhang_unreasonable_2018}} as a function of the number of scanned lines for various action selection policies. Every line represents a subject, where a dashed line uses harmonic imaging, solid lines use fundamental imaging}\wtwo{, and brackets indicate $p<0.05$ using the Wilcoxon signed-rank test.}}
    \label{fig:image_quality_in_house}
\end{figure}

\subsubsection{Reconstruction quality}

\wes{\autoref{fig:image_quality_in_house} shows the reconstruction quality in terms of \ac{PSNR} \wtwo{and LPIPS \cite{zhang_unreasonable_2018}} for three subsampling rates. CASL outperforms random and equispaced undersampling for all subsampling rates and every subject in the dataset.}
The qualitative results are shown in \autoref{fig:inhouse}. 
\wes{Here, we see the focused transmits, diverging transmits, and reconstructions using \ois{CASL} for two subjects, where one uses the fundamental mode and the other uses harmonic imaging.}
Even though the diffusion model was trained on a different dataset, the method still reconstructs well using limited measurements. For the same number of transmits as diverging waves, it shows certain details, such as the valve, more clearly.

\subsection{3D echocardiograms} \label{sec:exp-3d}
In this section, we apply \ois{CASL} to 3D \on{echocardiography}. Following Stevens \textit{et al.}~\cite{stevens2025high}, we consider a measurement model in which the elevation dimension is sparsely sampled, leading to a small set of acquired focused elevation planes from which the full volume must be recovered. Building on the reconstruction model implemented by Stevens \textit{et al.}, we too train a \ac{DM} on 2D slices taken along the axial ($\text{ax}$) and elevation ($\text{el}$) axes, but we extend this model in the temporal direction as with our EchoNet model described in Section~\ref{sec:method}. Our prior is therefore approximating the joint distribution $p(\tX)$ where $\tX \in \mathbb{R}^{N_\text{ax} \times N_\text{el} \times W}$. \otwo{For this task, a new \ac{DM} was trained, to better match the resolution and distribution of $(\text{ax}, \text{el})$ slices, which depart from the apical four-chamber views used in the previous experiments.} The \ac{DM} was trained on samples of size $N_\text{ax}=400$, $N_\text{el}=48$, and $W=3$. The dataset consists of 100 \textit{in-vivo} \otwo{B-mode} volume sequences across 16 patients, acquired using a Philips EPIQ scanner with an X5-1c matrix probe \otwo{(Philips Research, North America). The transducer has a frequency range of 1 MHz to 5 MHz.} A set of 8 volume sequences across 3 patients is held out for testing. \otwo{The test volumes contained 400 axial samples, 48 elevation planes, and varying numbers of azimuthal angles, ranging from 56 to 84.} For posterior sampling, a guidance weight of $\gamma=3$ was used, with $N_p=2$, $\tau_\text{max}=500$, and $\tau_\texttt{SeqDiff}=450$, and initial planes $A_1$ selected uniformly at random.

In order to perform the action step on 3D volumes, the K-Greedy Entropy Minimization algorithm was modified to first average the entropy map across azimuthal angles to produce a 2D entropy map along the axial and elevation axes. Given this 2D entropy map, the algorithm proceeds as in the 2D case, selecting a series of lines, now representing elevation planes, aiming to cover as much entropy as possible. 

\begin{figure}
    \centering
    \includegraphics[width=\linewidth]{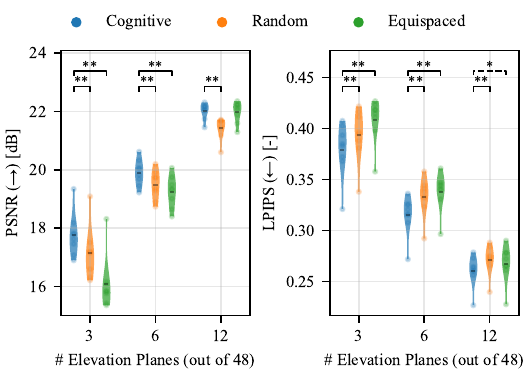}
    \vspace{-9pt}
    \caption{Reconstruction performance for the 3D dataset in terms of \acs*{PSNR} and \acs*{LPIPS} \wes{\cite{zhang_unreasonable_2018}} as a function of the number of scanned lines for various action selection policies. The figure shows a distribution over the data samples and includes the mean as a gray line \wtwo{(**$p<0.01$, *$p<0.05$ using the Wilcoxon signed-rank test).}}
    \label{fig:3d-psnr}
\end{figure}

\begin{figure}
    \centering
    \includegraphics[width=\linewidth]{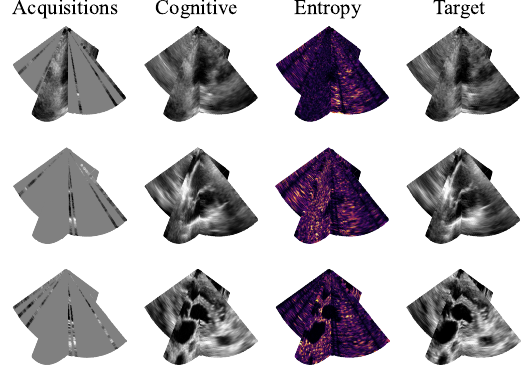}
    \vspace{-9pt}
    \caption{Qualitative results on the 3D dataset. The figure shows the acquisitions and reconstructions for 6 / 48 elevation planes compared to the target. Additionally, it shows the posterior entropy, which drives action selection.}
    \label{fig:3d-qualitative}
\end{figure}

As with the experiments on EchoNet-Dynamic, we benchmarked reconstructions created with \ois{CASL} against those created with baseline sampling strategies, with \ac{PSNR} and \ac{LPIPS}~\wes{\cite{zhang_unreasonable_2018}} results plotted in \autoref{fig:3d-psnr}. \otwo{The test volume sequences also varied in length, from 6 to 39 volumes long, and so the metrics were averaged per sequence to avoid bias}. Across the subsampling rates, it is clear that employing \ois{CASL} results in more faithful reconstructions, particularly with more aggressive subsampling. 
In \autoref{fig:3d-qualitative}, we provide qualitative examples in the form of bi-plane plots of volume reconstructions from 6/48 elevation planes, at the $4^{th}$ frame in each sequence.

\section{Discussion}
It is clear throughout the results provided in Section~\ref{sec:experiments} that \ois{CASL} outperforms the equispaced and random baseline strategies. \on{The degree of improvement varies across the experiments.} In Section~\ref{sec:echonet-dynamic}, our results on the 2D EchoNet-Dynamic dataset show significant benefits to using \wes{CASL}, achieving an improved trade-off between reconstruction quality and subsampling rate. \moved{\on{In our experiments on 3D data in Section~\ref{sec:exp-3d}, we also find that \ois{CASL} outperforms fixed sampling strategies across a range of sampling budgets, achieving a better trade-off between volume rate and reconstruction accuracy \ois{than equispaced and random strategies}.}}

\moved{\otwo{It is noteworthy that CASL achieves strong performance on the in-house echocardiograms using measurements from both fundamental and harmonic imaging modes, while using a generative prior trained on EchoNet-Dynamic. This showcases an instance of \textit{generalization}, wherein the model recovers a target sample from a new, unseen distribution. Indeed, Jia \textit{et al.} have shown that, if enough measurements have been acquired, \ac{DPS} can recover target images from distributions that are vastly different from the prior \cite{jia2026weak}. In this case, the in-house echocardiograms have in common with EchoNet-Dynamic that they use the apical four-chamber view. To explore the limits of generalization with CASL, we have included additional results using measurements from parasternal long-axis echocardiograms, and a static CIRS phantom on the GitHub repository.}}

\subsection{Future work}

\otwo{Future work towards improving performance in the 2D regime might develop approaches to generative modeling that can support longer context windows, more efficient inference, and higher spatial resolution. For example, we now use a data prior trained on the EchoNet-Dynamic dataset, which consists of 112$\times$112 pixels, for our in-house dataset, which has the potential to be beamformed at a much higher resolution. The algorithm could also be improved by lowering reconstruction error. While it was shown in Section \ref{sec:experiments} that the reconstruction error did not significantly hamper performance on a left ventricle segmentation task, the impact of the specific character of reconstruction errors introduced by CASL on further downstream tasks could be interesting in future work.}

\wthree{Our encouraging 3D results} highlight opportunities for further enhancement. In particular, training on a substantially larger 3D dataset (e.g., millions of volumes) would likely improve the model’s reconstruction quality and the informativeness of our derived uncertainty estimates. Furthermore, focusing in both the elevation and azimuthal directions\wes{, as compared to elevation plane selection,} would significantly enlarge the action space and allow for more targeted, information-efficient acquisition. Together, these enhancements have the potential to significantly boost the effectiveness of \ois{cognitive subsampling} in 3D ultrasound.

In our experiment using in-house echocardiograms, we chose line-by-line beamforming, although \ac{RTBF} could potentially yield higher-quality images. However, with \ac{RTBF}, the measurement model $p(\by \mid \bx, A^\ell)$ becomes more challenging and no longer corresponds to an inpainting task. \wes{The same holds for other types of \wtwo{spatially overlapping reconstructions}, such as \wtwo{is common for} diverging waves or plane waves.} Future work could explore how to better leverage the image quality benefits of \ac{RTBF}. 

\wtwo{In our experiments we retrospectively subsample the selected lines.} To fully leverage \wes{CASL}, the algorithm must operate in real-time with the frame acquisition\wtwo{, which could improve the results due to higher temporal correlation}. \wes{We found that with GPU hardware from 2023, our algorithm \wtwo{could} increase frame rate \wtwo{from 46 Hz to 58 Hz} through subsampling to 89/112 lines. In \autoref{tab:inference_speed}, we computed that 28 scan lines result in a physical frame acquisition time of 5.46 ms. Therefore, to achieve higher subsampling rates, and higher frame rates,} the algorithm still requires an approximate \wes{3}$\times$ speed-up. \ois{A final avenue for future work could involve exploring clinical advantages of the additional frame-rate offered by CASL.}

\section{Conclusion}
\label{sec:conclusion}

We proposed a patient-adaptive focused transmit \wes{and receive} scheme that reduces the number of acquisitions needed for a high-quality \wes{echocardiogram} by actively selecting the most informative measurements. Our method leverages posterior sampling with a temporal diffusion model and \ois{acquires} new measurements where the approximated posterior shows the most entropy. We have shown \otwo{that our method} outperforms baselines on the 2D EchoNet-Dynamic dataset\ois{, a 2D in-house dataset consisting of both fundamental and harmonic imaging modes, and} a 3D \ois{cardiac} dataset, especially in cases with very \ois{few} focused transmits. We have shown that \ois{using cognitive ultrasound} with focused transmits improves \ac{gCNR} compared to diverging waves with the same number of transmits. 
\wes{For 112 $\times$ 112 images, t}he method can be run in real-time at \wes{58} Hz on GPU accelerators from 2023.

\section*{Acknowledgment}

The authors would like to thank Danique de Bruijn for performing the ultrasound scans.

\printbibliography

\end{document}